\documentclass[final,3p]{elsarticle}

\usepackage{graphics}
\usepackage{graphicx}
\usepackage{epsfig}

\usepackage{amssymb}
\usepackage{amsthm}

\usepackage{amsmath} 

\begin{document}

\begin{frontmatter}



\title{State-space formulation of scalar Preisach hysteresis model  for rapid
computation in time domain}


\author{Michael Ruderman\corref{cor1}}
\ead{ruderman@vos.nagaokaut.ac.jp}

\address{Nagaoka University of Technology, Department of Electrical Engineering,
Nagaoka, 940-2188, Japan}



\begin{abstract}
A state-space formulation of classical scalar Preisach model
(CSPM) of hysteresis is proposed. The introduced state dynamics
and memory interface allow to use the state equation, which is
rapid in calculation, instead of the original Preisach equation.
The main benefit of the proposed modeling approach is the reduced
computational effort which requires only a single integration over
the instantaneous line segment in the Preisach plane. Numerical
evaluations of the computation time and model accuracy are
provided in comparison to the CSPM which is taken as a reference
model.
\end{abstract}


\end{frontmatter}


\section{Introduction}
\label{sec:1}

The Preisach-based modeling approaches
\cite{Preis35}--\nocite{Maye03}\nocite{Visit94}\cite{BertMayer06}
are widely used to describe the hysteresis effects of different
nature. The classical scalar Preisach model (CSPM) bases upon two
mathematical properties: wiping-out and congruency \cite{Maye03}.
Numerous extensions of CSPM \cite{BertMayer06} were proposed over
the last decades in order to overcome several discrepancies
between CSPM and hysteresis effects experimentally observable in
magnetic materials and structures. Among these are the
non-congruency of minor hysteresis loops, rate-dependency (or
dynamic generalization) of hysteresis, as well as stabilization
process (or accommodation) which violates the wiping-out
hysteresis property. The evolution of Preisach-type hysteresis
models has been illustrated e.g. in \cite{BottChiamChiarRep00}.
Further reformulations of CSPM such as DOK model
\cite{DellTorreOtiKadar90} and so-called modified scalar Preisach
model (MSPM) \cite{CardDellTorreTell00} were proposed for a fast
hysteresis computation and with regard to reversible components of
magnetization. A comparative study of DOK and MSPM scalar
hysteresis models is also given in \cite{AzzerCardFinoc04}.
Despite a continuous advancement in hysteresis modeling the CSPM
remains a still significant means for simulation and control tasks
shown in various applications, see e.g. \cite{DupreEtAl98},
\cite{NataleVelarVison01}, \cite{RosenEtAl10}.
It is worth noting that also another than Preisach formalisms,
e.g. Duhem \cite{GentiliGiorgio01} or first-order nonlinear
differential equations \cite{ChuaStrom71}, \cite{Visit94} have
been widely used for modeling hysteresis. For instance the
dissipativity property and stability of systems with the Duhem
hysteresis operator have been recently studied in
\cite{Jayaward12}. Further, the so-called butterfly-shaped
hysteresis map and its conversion to simple piecewise monotone
hysteresis maps has been analyzed in \cite{Drincig11}. The
significance of appropriate hysteresis modeling in control can be
seen for both cases: immediate use as an inverse filter for
compensating hysteresis \cite{Tan04}, and analyzing the effect of
hysteresis on the stability of control system to be designed.

When using a Preisach-type hysteresis modeling, in particular the
computational efficiency is required due to a large number of
hysteresis operators, which are spatially distributed in space of
the threshold values. Especially, a continuous (dynamic)
hysteresis behavior is computationally intensive since the smooth
input time series provoke continuous changes of the hysteresis
output. Here it should be noted that the smooth input time series
relate to the physical signals in various applications where the
input jumps and impact of noise are excluded from consideration,
e.g. simulation and/or feed-forwarding of controlled processes.

The purpose of this communication is to provide a state-space
formulation of CSPM in order to reduce the original computation
costs without losing significantly the model accuracy at certain
sampling conditions. The proposed modeling approach was first
shown at the International Symposium on Hysteresis and
Micromagnetics Modeling (HMM2011) and since then applied in the
discrete-time form in \cite{Rud12a,Rud13}. In the recent work, the
general continuous-time and continuous-state case is provided in
details. The main benefit is the computation effort reduced from
the double integration over the Preisach plane to a single
integration over the line segment at an operation state. In Landau
notation it means that the computation complexity of CSPM, which
is originally $\mathcal{O}$$(n^2 + 1)$, is reduced to
$\mathcal{O}$$(n + 1)$. Here, it is important to note that the
majority of Preisach-type hysteresis modeling approaches use the
so called first-order reversal curves (FORC), see e.g.
\cite{Maye03}, \cite{BertMayer06} for details, so as to avoid the
double integration. This widely elaborated and well-implementable
approach provides indeed a fast hysteresis computation, but, at
the same time, requires discretizing the Preisach plane in order
to store a finite set of FORCs. This does not allow applying the
continuous and parametric Preisach distribution functions and
limits the simulation capabilities to the discretized input-output
maps. In rest of the paper the details of the proposed model
formulation and various numerical evaluation examples are given.

\section{State-space formulation}
\label{sec:2}

\begin{figure}[!h]
\centering
\includegraphics[width=0.35\columnwidth]{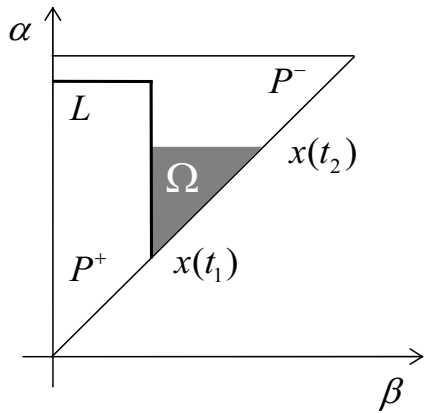}
\caption{Preisach plane with switching region} \label{fig:1}
\end{figure}

In order to provide the CSPM in state-space formulation let us
briefly summarize first the main CSPM equations and geometrical
interpretation by means of the Preisach plane (further denoted by
$P$). The basic input-output map of Preisach hysteresis operator
$\mathrm{H}[\cdot]$ is given by
\begin{equation}
y(t) = \mathrm{H} \, [x(t)] = \iint\limits_{\alpha \geq \beta}
\rho(\alpha,\beta) \, h_{\alpha\beta} [x(t)] \mathrm{d}\alpha
\mathrm{d}\beta \,. \label{eq:preisachoper}
\end{equation}
The elementary hysteresis operator (\emph{hysteron})
$h_{\alpha\beta}[\cdot]$ constitutes a non-ideal (delayed) relay
with 'up' and 'down' threshold values $\alpha$ and $\beta$. The
main parametrization factor of CSPM is the Preisach density
function (PDF) $\rho(\alpha,\beta)$ defined over $P  =
\{(\alpha,\beta) \; | \; \alpha \geq \beta\}$. At each instant $t$
the plane $P$ is divided into two disjunct subsets, $P^{+}(t)$
which contains the hysterons in the 'up' state ($+1$) and
$P^{-}(t)$ which contains the residual hysterons in the 'down'
state ($-1$). Both subsets are separated by a staircase interface
$L$, as illustrated schematically in Figure \ref{fig:1}, which
constitutes the instantaneous memory state of hysteresis.
According to $h_{\alpha\beta}: x \mapsto \{-1,+1\}$ and with
respect to $P(t)=P^{+}(t) \cup P^{-}(t)$ one can easily obtain
\begin{eqnarray}
\nonumber y(t) = & \iint\limits_{P_{+}} \rho(\alpha,\beta)
\textrm{d}\alpha \textrm{d}\beta -
\iint\limits_{P_{-}} \rho(\alpha,\beta)  \textrm{d}\alpha \textrm{d}\beta = \\
 = & 2  \iint\limits_{P_{+}} \rho(\alpha,\beta) \textrm{d}\alpha
\textrm{d}\beta - \iint\limits_{P} \rho(\alpha,\beta)
\textrm{d}\alpha \textrm{d}\beta \: . \label{eq:preisachout}
\end{eqnarray}
When increasing the input, starting from $x(t_1)$ and going
monotonically to $x(t_2)$ with $t_2 > t_1$, the hysterons within
the switching region $\Omega$ transit to the 'up' state, thus
changing from $P^{-}$ to $P^{+}$. The output difference $\Delta y
= y(t_{2}) - y(t_{1})$ can be obtained from Eq.
(\ref{eq:preisachout}) as
\begin{eqnarray}
\nonumber \Delta y = & 2  \iint\limits_{P_{+}(t_{2})}
\rho(\alpha,\beta) \textrm{d}\alpha \textrm{d}\beta & - \quad 2
\iint\limits_{P_{+}(t_{1})} \rho(\alpha,\beta)  \textrm{d}\alpha \textrm{d}\beta = \\
= & 2 \iint\limits_{\Omega} \rho(\alpha,\beta) \, \textrm{d}\alpha
\textrm{d}\beta \:, &  \label{eq:preisachoutdiff}
\end{eqnarray}
this with respect to the set definition $P^{+}(t_{2})=P^{+}(t_{1})
\cup \Omega$. Since the Preisach hysteresis function is monotonic
and the variable switching region captures the output difference
for all increasing and decreasing branches one can write the
general form of the output difference as
\begin{equation}
\Delta y  = 2 \mathrm{sgn} (\Delta x) \iint\limits_{\Omega(\Delta
x)} \rho(\alpha,\beta) \textrm{d}\alpha \textrm{d}\beta \: .
\label{eq:preisachdiff}
\end{equation}
Equation (\ref{eq:preisachdiff}) is valid for all input
differences $\Delta x = x_2 - x_1$ at finite time $\Delta t = t_2
- t_1 > 0$. Note that CSPM constitutes a rate-independent
hysteresis map so that $\Delta y$ depends on $\Delta x$ only and
not on $\Delta t$.

Now consider the continuous nonlinear model
\begin{equation}
\frac{\mathrm{d}\mathbf{z}}{\mathrm{d} t} = \mathbf{f} \Bigl(
\mathbf{z}, \frac{\mathrm{d}x}{\mathrm{d} t} \Bigr)
\label{eq:ssgeneral}
\end{equation}
with the state vector $\mathbf{z}$ which contains the output value
$y(t)$ and memory vector $\mathbf{m}(t) = m \cup M$. The latter is
an ordered set of alternating local minima
\begin{equation}
m \subset \bigl\{ x(t) \; | \; \dot{x}(t) = 0 \wedge x(t_{+})
> x(t) \wedge t_0 < t < t_{+} \bigr\} \label{eq:minima}
\end{equation}
and maxima
\begin{equation}
M \subset \bigl\{ x(t) \; | \; \dot{x}(t) = 0 \wedge x(t_{+}) <
x(t) \wedge t_0 < t < t_{+} \bigr\} \:. \label{eq:maxima}
\end{equation}
Recall that only the local extrema, which fulfill the wiping-out
hysteresis property \cite{Maye03}, are stored in the memory vector
and constitute the $(\alpha,\beta)$ vertexes of the interface $L$
between $P^{+}$ and $P^{-}$. That is each novel $m$ item,
according to Eq. (\ref{eq:minima}), is not only included to
$\mathbf{m}(t)$ but also erases from $\mathbf{m}(t)$ all the
previously stored minima which are less than the recent one. The
same mechanism is acting for each novel $M$ item which is larger
than the previously stored local maxima. Since the ground state
(or ``demagnetized state'' in context of magnetism \cite{Maye03})
of hysteresis is determined by the main diagonal $\alpha = -\beta$
(in the Preisach plane) the initial states (at $t=0$) are
\begin{equation}
y_0 = 0\, \quad \hbox{and} \quad \mathbf{m}_0 =
\left[%
\begin{array}{cc}
  x_{max} & x_{min} \\
  0 & 0 \\
\end{array}%
\right] \:. \label{eq:initialstate}
\end{equation}
Note that the $(x_{max},x_{min})$ vertex, which is the opposite
end of $L$ relating to the operation point
$\bigl(x(t),x(t)\bigr)$, is required to margin the Preisach plane
in the implementation.

When taking the limiting value $\Delta x \rightarrow 0$ and
dividing the left- and right-hand sides of Eq.
(\ref{eq:preisachdiff}) by the time derivative the output state
equation is represented by
\begin{equation}
\frac{\mathrm{d}y}{\mathrm{d}t}  = 2 \, T \, \mathrm{sgn}
(\mathrm{d}x) \iint\limits_{\Omega(\mathrm{d}x)}
\rho(\alpha,\beta) \textrm{d}\alpha \textrm{d}\beta \: .
\label{eq:state1}
\end{equation}
Note that the state equation (\ref{eq:state1}) complies with the
general one given in Eq. (\ref{eq:ssgeneral}), since $\Omega$ is
obtained using the instantaneous memory state and $\mathrm{d}x$
value. The latter is available for a constant computation rate,
i.e. $\mathrm{d}t = \mathrm{const} = 1/T$.

\begin{figure}[!h]
\footnotesize
\centering (a) \hspace{4.5cm} (b) \\[0.3cm]
\includegraphics[width=0.3\columnwidth]{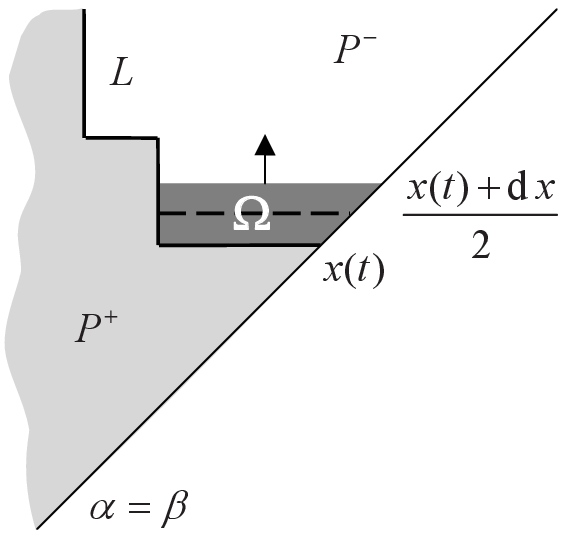}\hspace{1cm}
\includegraphics[width=0.3\columnwidth]{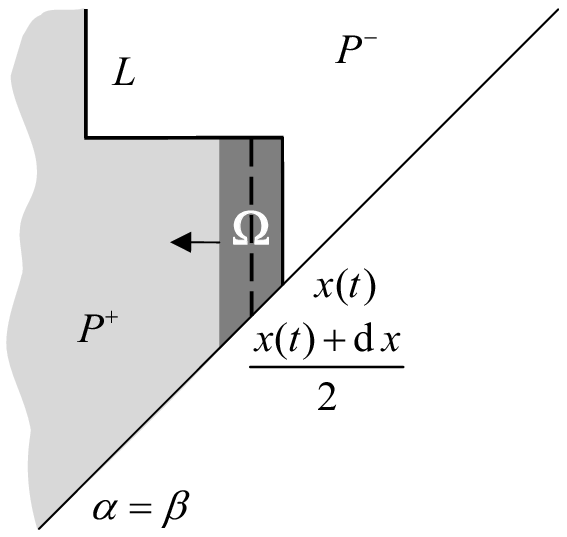}
\caption{Integral area by increasing (a) and decreasing (b) input}
\label{fig:2}
\end{figure}

In order to integrate PDF over the variable switching region
consider an arbitrary hysteresis state, once for increasing and
once for decreasing input as exemplarily shown in Figure
\ref{fig:2}.
\normalsize Since $\Delta x \rightarrow 0$ the switching region
converges to a line segment with
$\alpha=\bigl(x(t)+\mathrm{d}x\bigr)/2$ for increasing, and
$\beta=\bigl(x(t)+\mathrm{d}x\bigr)/2$ for decreasing input value.
Hear, the opposite end of the line segment is determined by the
latest stored minima ($m_l$) for $\mathrm{d}x > 0$ and maxima
($M_l$) for $\mathrm{d}x < 0$ correspondingly. Hence, the double
integration over the switching region can be replaced by the
single integration over the line segment when $|\mathrm{d}x|$ is
small enough. Denoting the $\bigl(x(t)+\mathrm{d}x\bigr)/2$ value
by $x_{+} \: \forall \: \mathrm{d}x > 0$ and by $x_{-} \: \forall
\: \mathrm{d}x < 0$ the state equation of hysteresis output can be
rewritten as
\begin{equation}
f_1 = \left\{%
\begin{array}{ll}
2 \, T \, \mathrm{sgn} (\mathrm{d}x) \int\limits_{\beta(m_{l})}^{x_{+}} \rho \bigl(x_{+},\beta \bigr)  \, \textrm{d}\beta , & \hbox{if } \;  \mathrm{d}x > 0 \\[0.3cm]
2 \, T \, \mathrm{sgn} (\mathrm{d}x) \int\limits_{\alpha(M_{l})}^{x_{-}} \rho \bigl(\alpha,x_{-} \bigr)  \, \textrm{d}\alpha , & \hbox{if } \;  \mathrm{d}x < 0\\[0.6cm]
    0, & \hbox{otherwise} \; .
\end{array}%
\right.  \label{eq:ssf1}
\end{equation}

Since the hysteresis memory state is described by a set of
$(\alpha,\beta)$ vertexes the memory state equation
$\mathrm{d}\mathbf{m}/\mathrm{d}t = f_2(\mathbf{z},
\mathrm{d}x/\mathrm{d}t)$ should be introduced by means of a
dedicated operator
\begin{equation}
f_2 = \left\{%
\begin{array}{ll}
    O_1^{+} \wedge O_2^+ \wedge O_3^+, & \hbox{if }   \;  \mathrm{d}x > 0    \\[0.2cm]
    O_1^{-} \wedge O_2^- \wedge O_3^-, & \hbox{if }   \;  \mathrm{d}x < 0    \\[0.2cm]
    \emptyset                          & \hbox{otherwiese} \:,\\
\end{array}%
\right. \label{eq:ssf2}
\end{equation}
using the algebra of sets. Note that the operator $f_2$ is not
commutative concerning $O_i$, because of the ordered set of the
memory vector. The single set operators
\begin{equation}
O_{1,\ldots,3}^+ =
\left\{%
\begin{array}{l}
    \backslash \: \bigl\{ m, M \: | \: \alpha \leq x_{+} \bigr\}  \\[0.2cm]
    \cup       \: \bigl(x_{+}, \, \beta(M_l)\bigr)  \\[0.2cm]
    \cup       \: \bigl(x_{+}, \, x_{+}\bigr) \\
\end{array}%
\right.
 \label{eq:setop1}
\end{equation}
and
\begin{equation}
O_{1,\ldots,3}^- =
\left\{%
\begin{array}{l}
    \backslash \: \bigl\{ m, M \: | \: \beta \geq x_{-} \bigr\}  \\[0.2cm]
    \cup       \: \bigl(\alpha(m_l), \, x_{-}\bigr)  \\[0.2cm]
    \cup       \: \bigl(x_{-}, \, x_{-}\bigr) \\
\end{array}%
\right.
 \label{eq:setop2}
\end{equation}
modify the memory state vector at each instant $t$, where the last
case in Eq. (\ref{eq:ssf2}) denotes no changes of $\mathbf{m}$
when $\mathrm{d}x=0$. The first case in Eqs. (\ref{eq:setop1}) and
(\ref{eq:setop2}) constitutes the wiping-out hysteresis property,
and the second and third cases correspondingly update the last
local extremum and insert the actual operation point.

The introduced state equations capture the entire hysteresis
behavior independent of the initial state $[y_0, \mathbf{m}_0]$
which is assumed to be known. Note that when computing the output
value
\begin{equation}
y(t)  = 2 T \int\limits_{0}^{t} \mathrm{sgn}
\bigl(\mathrm{d}x(t)\bigr)
\iint\limits_{\Omega\bigl(\mathrm{d}x(t)\bigr)} \rho(\alpha,\beta)
\textrm{d}\alpha \textrm{d}\beta \: \textrm{d}t
 \label{eq:statey}
\end{equation}
the state equation (\ref{eq:ssf2}) has to be evaluated first
before integrating the output dynamics.

\section{Numerical evaluation}
\label{sec:3}

The proposed scalar Preisach model in a state-space formulation
(further as SSPM) is implemented in a dedicated software code
working within MATLAB$^\copyright$ computation environment. All
simulation results below are obtained on an ordinary PC powered by
an Intel Duo CPU, 2 GHz, with 4 GB random access memory (RAM). The
numerical integration is performed using the standard routine with
adaptive Lobatto quadrature, where the set error tolerance
(\emph{tol}) constitutes the main parametrization factor which
impacts both the integration accuracy and computation time.
\begin{figure}[!h]
\centering
\includegraphics[width=0.49\columnwidth]{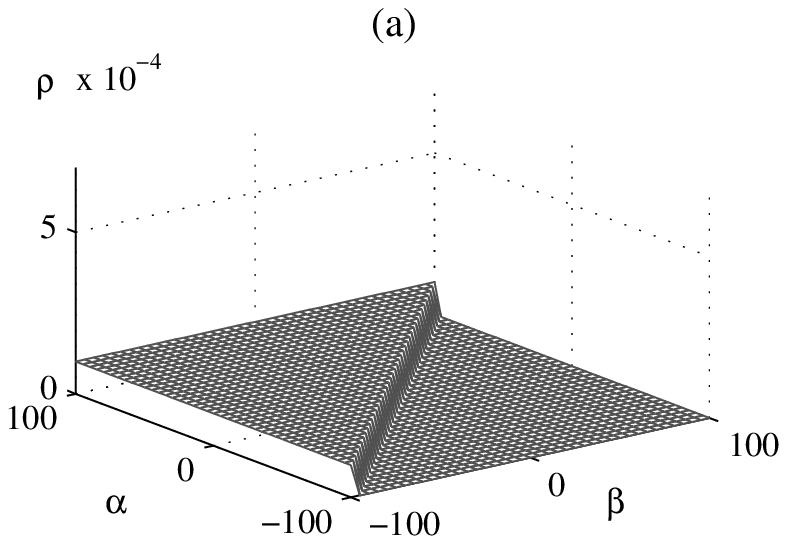}
\includegraphics[width=0.49\columnwidth]{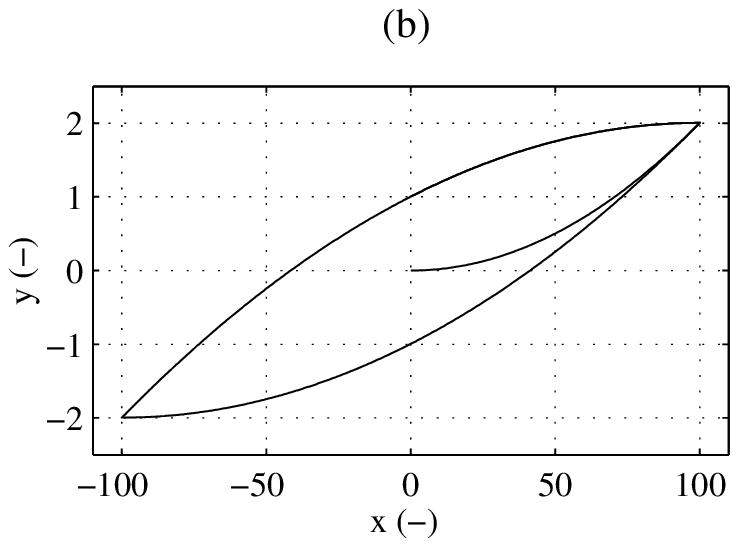}\vspace{0.3cm}
\includegraphics[width=0.49\columnwidth]{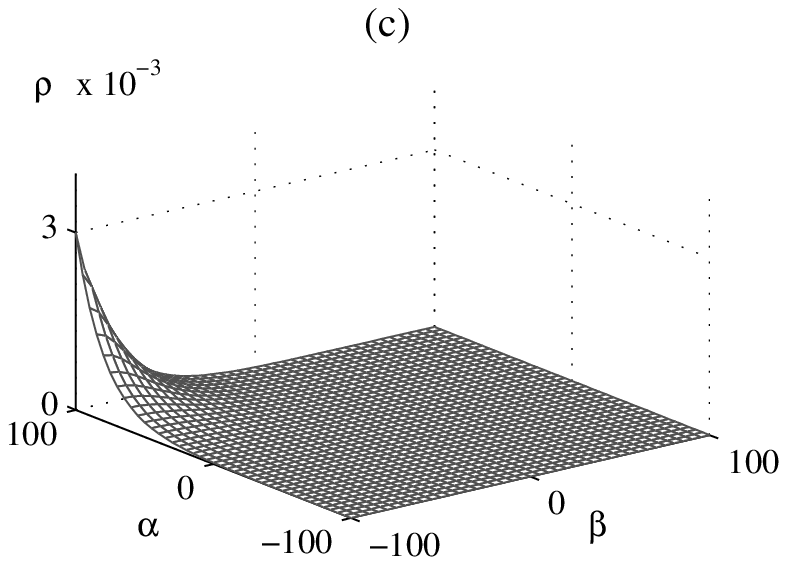}
\includegraphics[width=0.49\columnwidth]{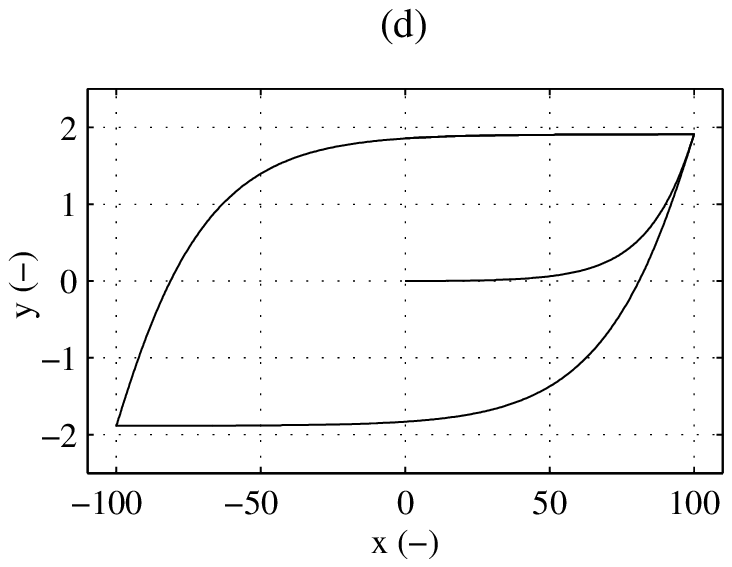}\vspace{0.3cm}
\includegraphics[width=0.49\columnwidth]{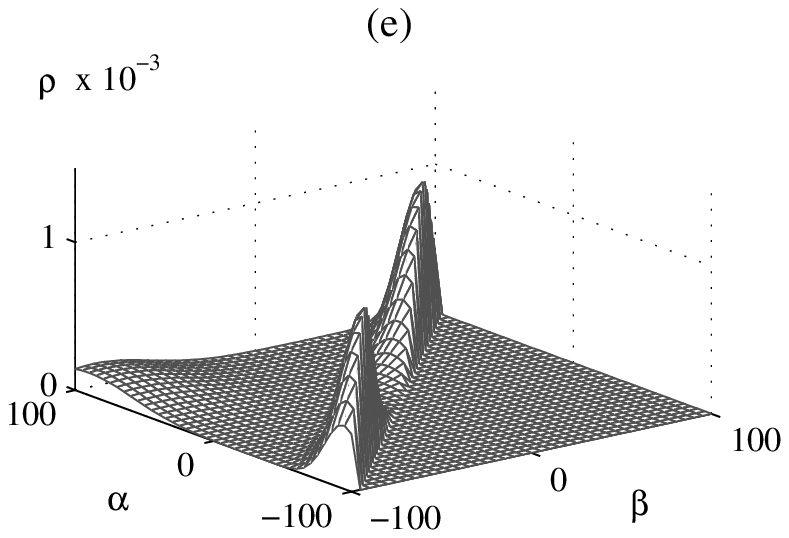}
\includegraphics[width=0.49\columnwidth]{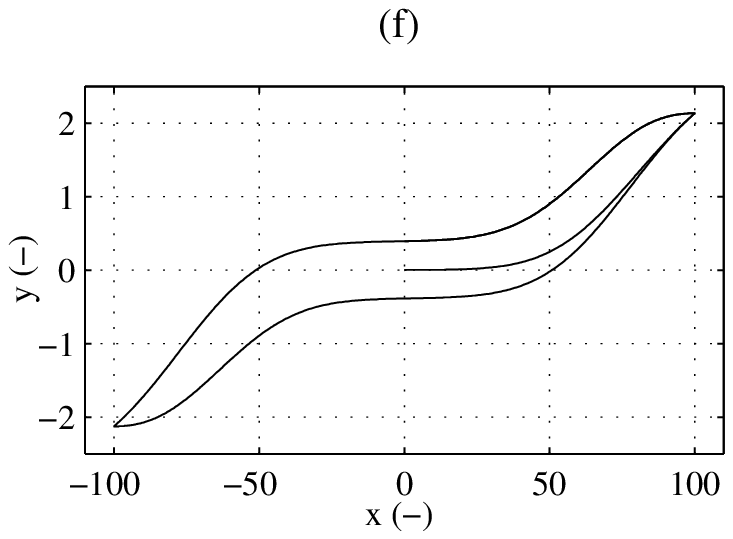}
\caption{Parametric PDFs and corresponding hysteresis loops:
uninform (a-b); Gaussian mixture (c-d); biased exponential (e-f)}
\label{fig:3}
\end{figure}

Three different parametric PDFs and that the uninform distribution
(abbreviated by U), biased exponential distribution (abbreviated
by E), and Gaussian mixture of three weighted normal distributions
(abbreviated by G), are taken for the model evaluation. The U-PDF
of level $v$ is given by
$$
\rho^U(\alpha,\beta) = v.
$$
The E-PDF with free parameters $A$, $B$, and $C$ is captured by
$$
\rho^E(\alpha,\beta) = A  \exp (-B  ||(\alpha,\beta)||) + C.
$$
The G-PDF with Gaussian components $\mathcal{N}(\mu, \Sigma)$,
each one parameterized by the mean vector $\mu$ and symmetric
covariance matrix $\Sigma$ (both in the $(\alpha,\beta)$
coordinates) and weighted by $w$, is given by
$$
\rho^G(\alpha,\beta) = \sum_{i=1}^{3} w_i \,  \mathcal{N}(\mu_i,
\Sigma_i)[\alpha,\beta].
$$
The assumed $(\alpha,\beta)$ distributions and the corresponding
major hysteresis loops started from the ground state are depicted
in Figure \ref{fig:3}. Note that the number of parameters of U-,
E-, and G-PDFs accounts for 1, 3, and 18 correspondingly.

\subsection{Comparison of CSPM and SSPM}
\label{sec:3:1}

First, the computation time and accuracy of SSPM are compared with
those achieved by CSPM as it is provided in Eq.
(\ref{eq:preisachoper}). Three error tolerance values
(\emph{tol}), $1\mathrm{e}^{-3}$, $1\mathrm{e}^{-4}$, and
$1\mathrm{e}^{-5}$ set for numerical integration are taken into
evaluation. The maximal $\hat{\tau}$ and average $\bar{\tau}$
computation time required to determine one closed (major)
hysteresis loop, as well as its standard deviation $\tilde{\tau}$,
are taken as performance metrics (\emph{pm}) of the computation
effort. In order to evaluate the model accuracy the CSPM response
with $1\mathrm{e}^{-5}$ \emph{tol} value is taken as reference
(\emph{ref}) since that one provides a most exact, but also most
time-consuming, integration of PDF. The accuracy performance
metrics -- maximal $\hat{e}$ and average $\bar{e}$ error as well
as the error standard deviation $\tilde{e}$ -- rely on the
relative error normalized by the overall hysteresis magnitude as
\begin{equation}
e\bigl(x(t)\bigr) = \frac{|y(t)-y_{ref}(t)|}{ |y_{max}-y_{min}|}
\times 100 \% \:.
 \label{eq:errorrel}
\end{equation}

One closed (major) hysteresis loop is computed using the periodic
input $x(t)=100 \sin(2 \pi \, t - \pi/2)$ and having the
simulation sampling time 0.001 sec. Thus, 1000 time-equidistant
$(x,y)$ data tuples per hysteresis loop are recorded for
evaluation. The defined performance metrics of computation time
and accuracy are compared in Table \ref{tab:SSPMcompCSPM} for CSPM
and SSPM. Note that the computation time metrics are denoted in
milliseconds (ms), and accuracy metrics are denoted in percent
(\%).
\begin{table}[ht]
  \renewcommand{\arraystretch}{1.5}
  \caption{Computation time and accuracy metrics of CSPM and SSPM}
  \label{tab:SSPMcompCSPM}
  \scriptsize
  \begin{center}
  \begin{tabular} {|p{0.8cm}|p{0.8cm}|p{0.8cm}||p{0.8cm}|p{0.8cm}||p{0.8cm}|p{0.8cm}||p{0.8cm}|p{0.8cm}|}
  \hline
  \multicolumn{3}{| c ||}{case}                      & \multicolumn{2}{c||} {U-PDF} & \multicolumn{2}{c||} {E-PDF} &  \multicolumn{2}{c|} {G-PDF} \\
  \hline
  \emph{tol}       & \emph{pm}            &  unit    & CSPM       & SSPM & CSPM       & SSPM & CSPM       & SSPM \\
  \hline \hline
                   & $\hat{\tau}$         &  ms      & 100        & 0.70 & 18.3       & 1.70 & 195        & 9.30 \\
  \cline{2-9}
                   & $\bar{\tau}$         &  ms      & 49.9       & 0.33 & 9.30       & 0.35 & 48.1       & 4.20 \\
  \cline{2-9}
  1e$^{-3}$        & $\tilde{\tau}$       &  ms      & 36.7       & 0.26 & 4.80       & 0.14 & 48.2       & 2.30 \\
  \cline{2-9}
                   & $\hat{e}$            &  \%      & \emph{1.22}       & \emph{2.80} & \emph{0.57}       & \emph{0.22} & \emph{4.89}       & \emph{2.15} \\
  \cline{2-9}
                   & $\bar{e}$            &  \%      & \emph{0.32}       & \emph{1.62} & \emph{0.23}       & \emph{0.18} & \emph{2.00}       & \emph{1.09} \\
  \cline{2-9}
                   & $\tilde{e}$          &  \%      & \emph{0.38}       & \emph{0.90} & \emph{0.24}       & \emph{0.18} & \emph{2.25}       & \emph{1.15} \\
  \hline \hline
                   & $\hat{\tau}$         &  ms      & 349        & 1.30 & 58.8       & 2.40 & 3600       & 14.2 \\
  \cline{2-9}
                   & $\bar{\tau}$         &  ms      & 165        & 0.72 & 13.5       & 0.45 & 499        & 7.30 \\
  \cline{2-9}
  1e$^{-4}$        & $\tilde{\tau}$       &  ms      & 129        & 0.32 & 10.6       & 0.25 & 588        & 3.40 \\
  \cline{2-9}
                   & $\hat{e}$            &  \%      & \emph{0.34}       & \emph{0.58} & \emph{0.21}       & \emph{0.12} & \emph{0.96}       & \emph{0.58} \\
  \cline{2-9}
                   & $\bar{e}$            &  \%      & \emph{0.17}       & \emph{0.40} & \emph{0.03}       & \emph{0.05} & \emph{0.19}       & \emph{0.33} \\
  \cline{2-9}
                   & $\tilde{e}$          &  \%      & \emph{0.19}       & \emph{0.13} & \emph{0.04}       & \emph{0.04} & \emph{0.24}       & \emph{0.13} \\
  \hline \hline
                   & $\hat{\tau}$         &  ms      & 495        & 3.00 & 172        & 2.30 & 4446       & 20.2 \\
  \cline{2-9}
                   & $\bar{\tau}$         &  ms      & 210        & 1.30 & 39.0       & 0.55 & 721        & 12.6 \\
  \cline{2-9}
  1e$^{-5}$        & $\tilde{\tau}$       &  ms      & 164        & 0.48 & 27.7       & 0.39 & 791        & 3.70 \\
  \cline{2-9}
                   & $\hat{e}$            &  \%      & \emph{ref} & \emph{0.03} & \emph{ref} & \emph{0.12} & \emph{ref} & \emph{0.06} \\
  \cline{2-9}
                   & $\bar{e}$            &  \%      & \emph{ref} & \emph{0.01} & \emph{ref} & \emph{0.05} & \emph{ref} & \emph{0.02} \\
  \cline{2-9}
                   & $\tilde{e}$          &  \%      & \emph{ref} & \emph{0.01} & \emph{ref} & \emph{0.04} & \emph{ref} & \emph{0.02} \\
  \hline
  \end{tabular}
  \end{center}
\end{table}
\normalsize

It is evident that the \emph{tol} value 1e$^{-5}$ provides the
most accurate but also the slowest hysteresis computation, and
that for all considered PDFs. Here, the accuracy metrics of SSPM
fall under 0.1 \% which is reasonably an acceptable value,
particularly in regard to unavoidable data errors when identifying
the hysteresis. The computation time metrics of SSPM are
cardinally superior in comparison to CSPM. Almost all
$\tau$-metrics of SSPM are 100--200 times lower than these of
CSPM, excepting the E-PDF case with lower \emph{tol} values at
which the differences are about 10--20 times. Remarkable is the
fact that a decrease of \emph{tol} value from 1e$^{-3}$ to
1e$^{-5}$, i.e. of two orders, leads to an increase of
$\tau$-metrics in the range of 5--20 times for CSPM, where the
$\tau$-metrics of SSPM increase in the range of 1.5--3 times only.

\subsection{Integration accuracy of SSPM}
\label{sec:3:2}

Since the computation of SSPM output bases on a continuous
integration of dynamic state, the integration errors of a
cumulative character can significantly disturb the model accuracy.
Besides an inherent accuracy, determined by the set error
tolerance value, the chosen finite sampling time $\mathrm{d}t$ can
additionally influence the state values computed according to Eq.
(\ref{eq:state1}), while the equal input sequence is applied. It
is evident that a lower sampling time leads to a higher model
accuracy since this reduces $\mathrm{d}x$, thus capturing the
switching region by the corresponding line segment in a more
accurate way.
\begin{figure}[!h]
\centering
\includegraphics[width=0.49\columnwidth]{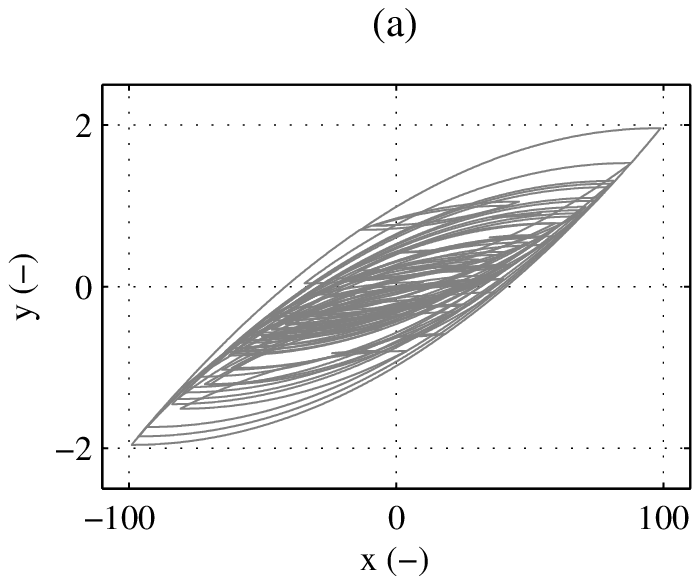}
\includegraphics[width=0.49\columnwidth]{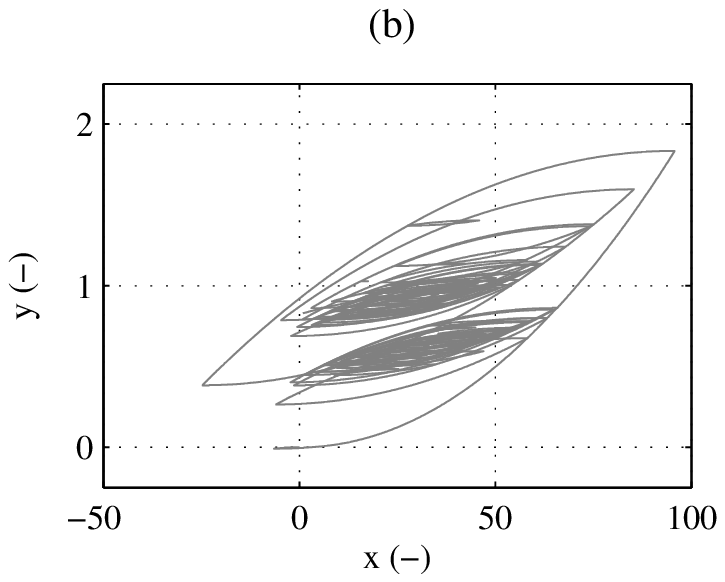}
\caption{Hysteresis response to the multi-sinus (0.1--10 Hz) input
sequence: (a) zero symmetrical, (b) zero biased} \label{fig:4}
\end{figure}
Two different multi-sinus input sequences with a bandwidth 0.1--10
Hz are evaluated on SSPM for the assumed U-PDF. The first one is
symmetrical to zero and the second one is biased, so that the
hysteresis operates predominantly in the first quadrant. The
resulting hysteresis response is visualized in Figure \ref{fig:4}
(a) and (b) correspondingly.
\begin{figure}[!h]
\centering
\includegraphics[width=0.49\columnwidth]{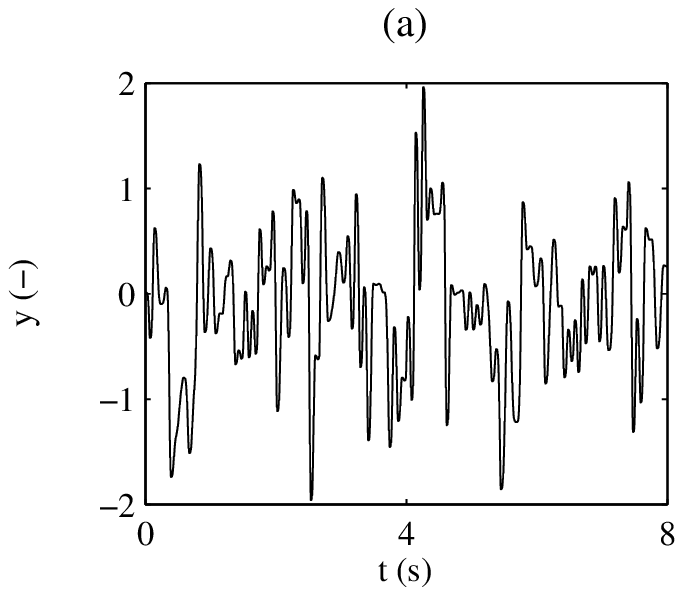}
\includegraphics[width=0.49\columnwidth]{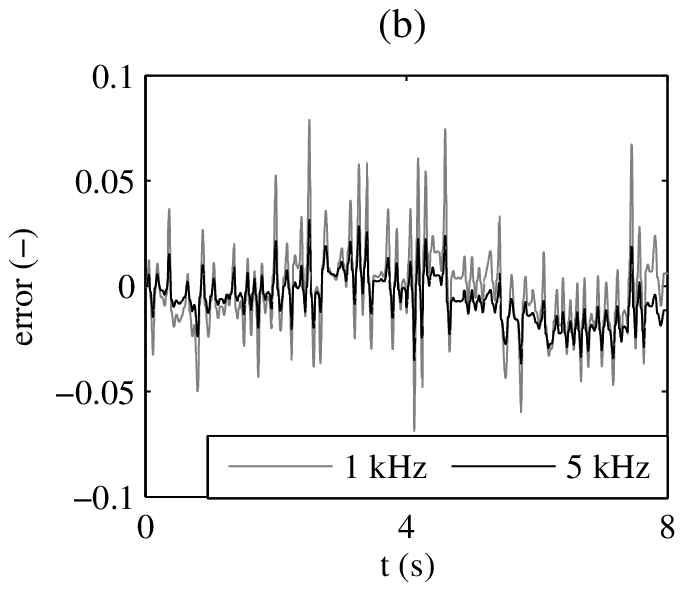}\vspace{0.15cm}
\includegraphics[width=0.49\columnwidth]{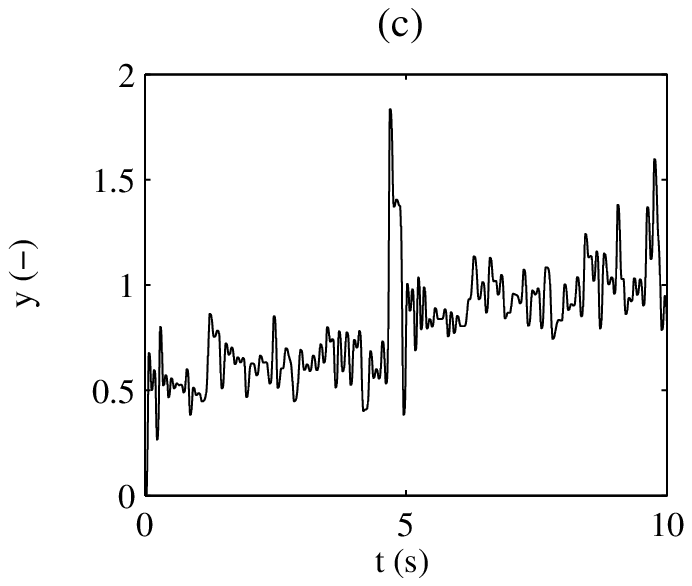}
\includegraphics[width=0.49\columnwidth]{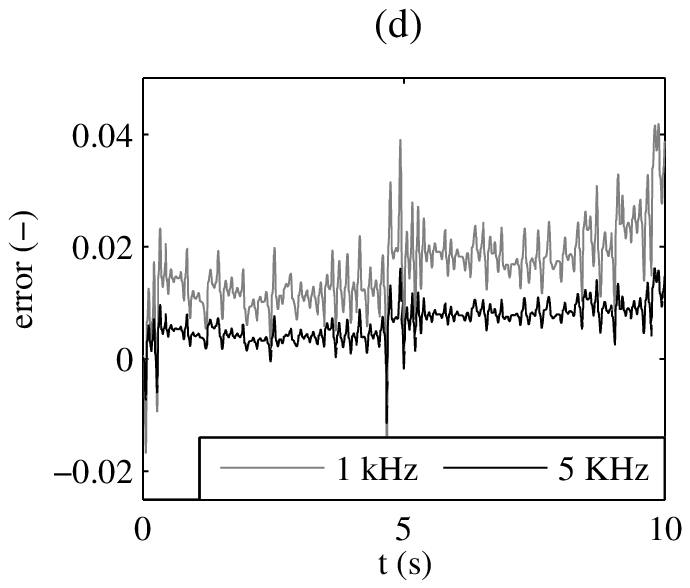}
\caption{Hysteresis output and evaluated relative error for
different sampling rates: (a-b) zero symmetrical, (c-d) zero
biased} \label{fig:5}
\end{figure}

In order to evaluate the integration accuracy of SSPM the sampling
rate of 10 kHz is taken as the reference. The reduced sampling
rates of 5 kHz and 1 kHz are evaluated in terms of the output
divergence comparing to the output obtained at 10 kHz. In Figure
\ref{fig:5} the hysteresis output and the evaluated relative error
are depicted for both multi-sinus sequences. It is evident that
the error obtained at 5 kHz sampling rate is lower than that
obtained at 1 kHz. Particularly, the fast higher peaks of
hysteresis input, and consequently output, lead to a higher
relative error. It comes as not surprising since the higher input
harmonics result in an increase of $\mathrm{d}x$, while
$\mathrm{d}t$ remains constant. Another important aspect is a
slight drift of the relative error observable in case of the
biased multi-sinus input. This appears since the DC signal
component provokes an integrating output error. However, the
latter can remain alternating around the DC bias when the total
number of ascending and descending hysteresis branches are well
balanced. A more extensive analysis of integration error dynamics
and evaluation of different numerical solver types could help
characterizing better the model accuracy in the future works.

\section{Conclusions}
\label{sec:4}

In this paper, a state-space formulation of scalar Preisach
hysteresis model has been introduced. By doing this the main
Preisach formalism and model properties, i.e. wiping-out and
congruency, are not violated. Using the same geometrical
interpretation by means of the Preisach plane and hysteresis
density function, which is the principal parametrization factor,
the dynamic state variables of hysteresis memory and output have
been introduced.

When using the Preisach hysteresis model in its original
formulation, the double integration of Preisach density function
over the operation space of hysterons is required. In this work,
the original computation cost have been reduced without losing
significantly the model accuracy. In Landau notation, the
computation complexity has been reduced from $\mathcal{O}$$(n^2 +
1)$ to $\mathcal{O}$$(n + 1)$, since the double integration over
the Preisach plane was replaced by the single integration over the
line segment determined by the operation state. Multiple numerical
evaluating tests have shown that the computation time performance
can be increased up to 100 times in comparison to the classical
scalar Preisach model. We note that the integration errors of
computing the dynamic state variable are the main sources of model
inaccuracies. However, reducing the error tolerance of numerical
integration and keeping an appropriate relation between the
sampling and input rate a fast and accurate model can be achieved.

Finally it can be concluded that the proposed state-space
formulation of classical scalar Preisach model is well suitable
for multiple hysteresis applications independent of the physical
nature and shape of congruent hysteresis loops. The modeled
hysteresis has, however, to comply with the assumed Preisach
hysteresis properties (see e.g. in \cite{Maye03}).





\bibliographystyle{elsarticle-num}
\bibliography{references}







\end{document}